\documentclass[a4paper,11pt]{article}
\pdfoutput=1 

\usepackage{jinstpub} 
\usepackage{graphicx}
\usepackage{subfigure}

\title{\boldmath A 500 MS/s waveform digitizer for PandaX dark matter experiments}
\author[a]{Changda He,}
\author[a,b]{Jianglai Liu,}
\author[c]{Xiangxiang Ren,}
\author[a]{Xiaofeng Shang,}
\author[a]{Xikai Wei,}
\author[a]{Mingxin Wang,}
\author[a]{Jijun Yang,}
\author[a]{Jinqun Yang,}
\author[a]{Yong Yang,} \note{Corresponding author.}
\author[d]{Guangping Zhang,}
\author[e]{Qibin Zheng}
\affiliation[a]{School of Physics and Astronomy, Shanghai Jiao Tong University, MOE Key Laboratory for Particle \\
  Astrophysics and Cosmology, Shanghai Key Laboratory for Particle Physics and Cosmology, 200 Dongchuan Road, Shanghai 200240, China}
\affiliation[b]{Tsung-Dao Lee Institute, 520 Shengrong Road, Shanghai 200240, China}
\affiliation[c]{Key Laboratory of Particle Physics and Particle Irradiation (MOE), Institute of Frontier and Interdisciplinary Science, Shandong University, 72 Binhai Road, Qingdao, Shandong 266237, China}
\affiliation[d]{College of Science, China University of Petroleum, 66 West Changjiang Road, Qingdao 266580, Shandong, China}
\affiliation[e]{Institute of Biomedical Engineering, University of Shanghai for Science and Technology, 516 Jungong Road, Shanghai 200093, China}

\emailAdd{yong.yang@sjtu.edu.cn}
\abstract{
  Waveform digitizers are key readout instruments in particle physics
  experiments. In this paper, we present a waveform digitizer for the
  PandaX dark matter experiments. It supports both
  external-trigger readout and triggerless readout, accommodating the
  needs of low rate full-waveform readout and channel-independent low
  threshold acquisition, respectively. This digitizer is a 8-channel
  VME board with a sampling rate of 500 MS/s and 14-bit resolution for
  each channel. A digitizer system consisting of 72 channels has been
  tested in situ of the PandaX-4T experiment. We report the system
  performance with real data.
}


\begin{document}
\maketitle
\flushbottom

\section{Introduction}
The dark matter (DM) makes up around 27\% of the Universe,
but its nature remains to be elusive. Direct detection and
identification of DM are among the most pressing challenges for
contemporary experimental physics. The PandaX project consists of a
series of experiments that use xenon as the detection media. These
experiments have produced most stringent limits on the couplings of some
well-motivated classes of DM to ordinary particles~\cite{pI_full,pII_99days,pII_SD,pII_54ton,pII_axion,pII_In,pII_EFT,pII_LMED,Cheng:2021fqb,pII_SIDM,p4paper}.

In PandaX DM experiments, the central apparatus is a dual-phase xenon
time projection chamber (TPC). Its cathode and gate are applied with
separate High Voltage (HV) to provide the drift electric field and
extraction field. Particle interacting with one of the xenon atoms in
the sensitive volume can generate a prompt flash of scintillation
lights (S1) and ionized electrons.  The latter are drifted up and
extracted into the gaseous region where they produce
electroluminescence, namely a second flash of scintillation lights
(S2). These lights are detected by two arrays of photomultipliers
(PMTs) instrumented at the top and the bottom of the TPC. The
electrical signals from PMTs are constantly sampled and digitized by
waveform digitizers.

The digitized samples constitute the most fundamental data for offline
analysis. For rare event search experiments like PandaX, all useful waveforms
should be read out, since they carry all information including the charge,
time, pulse shape and so on. However, not every sample can be read out. In the previous
experiments PandaX-I and PandaX-II, the waveform of each PMT is sampled
every 10 ns at a 14-bit resolution. So the data rate per digitizer
channel is 190 MB/s, which is already above the 85 MB/s maximum
readout bandwidth per digitizer.  Given the total number of channels
(for example, 158 in PandaX-II), the total bandwidth becomes extremely
large. Therefore, digitized samples must be read out selectively.

In PandaX DM experiments, two data acquisition (DAQ) schemes have been
developed, depending on chosen digitizers. In PandaX-I and
PandaX-II~\cite{Ren:2016ium, Wu:2017cjl}, the readout of the
digitizers (CAEN V1724, 100 MS/s sampling rate) relies on global
external trigger signals from a trigger system, which is designed
for capturing relatively large S2 signals. For each trigger, data from
all channels in a fixed time window around the trigger time can be
read out, with or without baseline suppressed depending on the need.
In order to save both S1 and S2 signals, the time window needs to
larger than twice of the maximum electron drift time. In the current
experiment PandaX-4T, another digitizer (CAEN V1725, 250MS/s) is used~\cite{p4daq}.
The readout does not require external triggers.  This is usually called
triggerless readout.  Each channel can be self-triggered and read out
independently with baseline suppressed.

The data acquired by external-trigger-based DAQ are affected by
inefficiencies of identifying S2 signals by the trigger system.  In
many DM searches, a lower cut on S2 signal is applied. For example, in
the PandaX-II light DM-electron analysis~\cite{Cheng:2021fqb}, the
lower cut is set to be the same as the trigger threshold, which refers
to the size of S2 signals when the trigger efficiency is 50\%~\cite{Wu:2017cjl}.
In this case, the S2 trigger inefficiency is one of the dominant causes for DM
detection efficiency loss. On the other hand, the triggerless DAQ is
designed to record all self-triggered data from all channels. S1 and
S2 signals are only identified offline.  S2 trigger inefficiency
becomes irrelevant.

However, triggerless readout does not meet all experimental needs.  In
this readout scheme, a readout buffer is usually needed to save
digitized samples, together with other information such as the trigger
time stamp, before they are finally read out to the DAQ server.  Due
to limited buffer space and readout bandwidth, the digitizers can
become busy due to fullness of the buffer.  In the commissioning runs
of PandaX-4T~\cite{p4daq}, the busy effect was found to be negligible
for the event reconstruction when the total date rates were less than
about 80 MB/s for dark matter runs and most calibration runs. In some
conditions with higher data rates, noticeable systematic errors were
observed in the event reconstruction~\cite{p4daq}. Therefore, it might
be more appropriate to use external-trigger-based scheme with baseline
suppression in these run conditions. In addition, to study the
baseline suppression effect in reconstructing S1 and S2 signals, it is
also desirable to read out full waveforms including both S1 and S2
signals during one trigger time window without baseline
suppressed. However, the chosen digitizers in PandaX-4T (V1725) cannot
record both S1 and S2 signals for each trigger with baseline
suppressed.
 
Besides the flexibility in readout schemes, a digitizer with higher
sampling rate is desired. Higher sampling rate is potentially
useful for improving the pulse shape discrimination in dual-phase
xenon TPC experiments~\cite{psd,wsw}. In addition, a new type of PMTs are
being considered for future PandaX experiment. These PMTs exhibit
shorter pulses for single photoelectron (SPE) signals compared to the
3-inch PMTs used in PandaX-4T. This also calls for a digitizer with a
higher sampling rate.
 
In this paper we present a 500 MS/s waveform digitizer designed to
support both external-trigger and triggerless readout schemes. A digitizer system consisting of
72 channels was built to acquire data in the PandaX-4T experiment in
June 2021, after its commissioning runs.  The in situ system
performance is presented. This digitizer has also been used to
evaluate the new PMTs. Its performance of measuring SPE signals is
presented.

\section{The Digitizer Design}
\label{hardware}

\begin{figure*}[!htbp]
  \centering
  \includegraphics[width=0.5\linewidth]{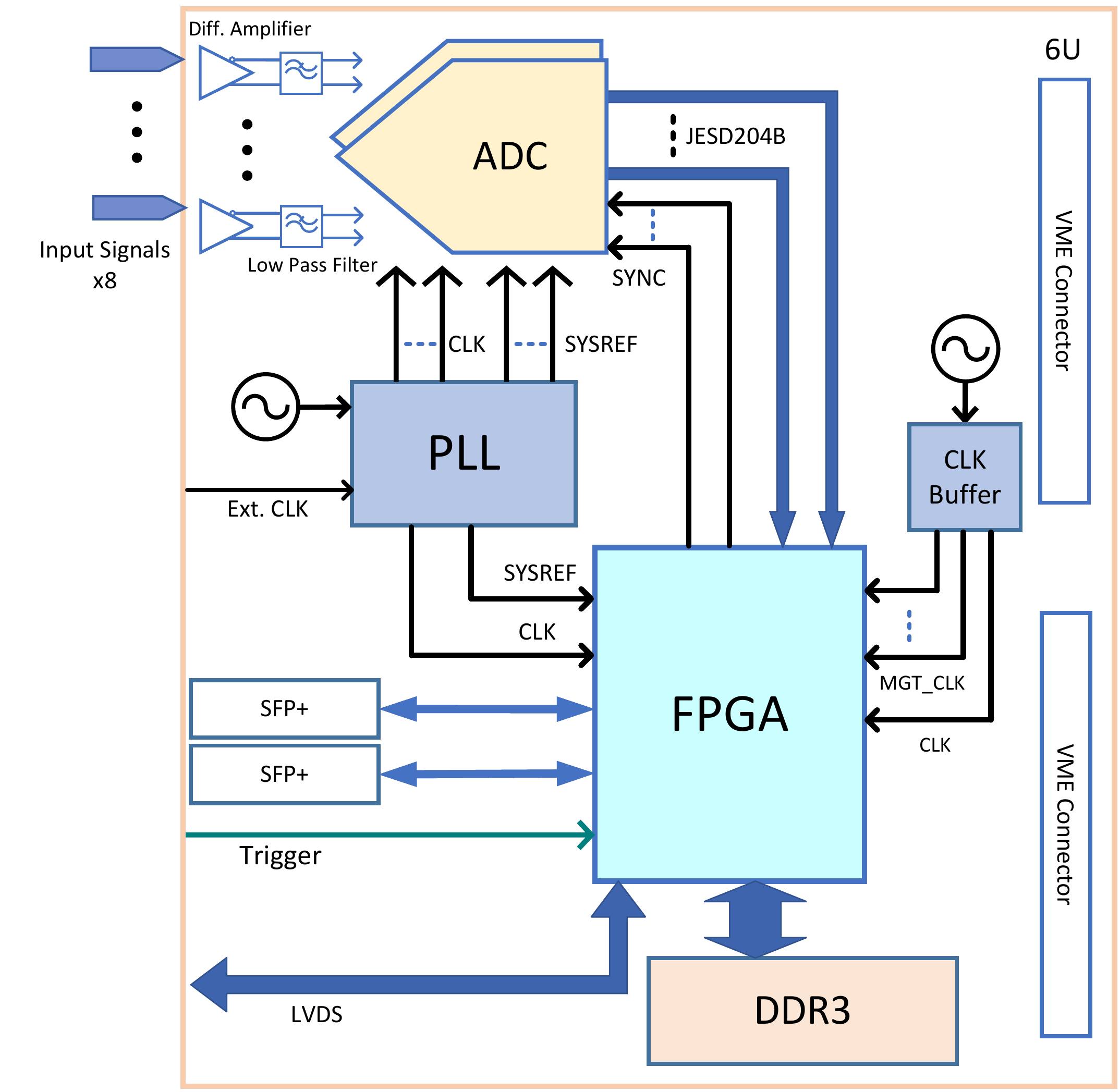}
  \includegraphics[width=0.4\linewidth]{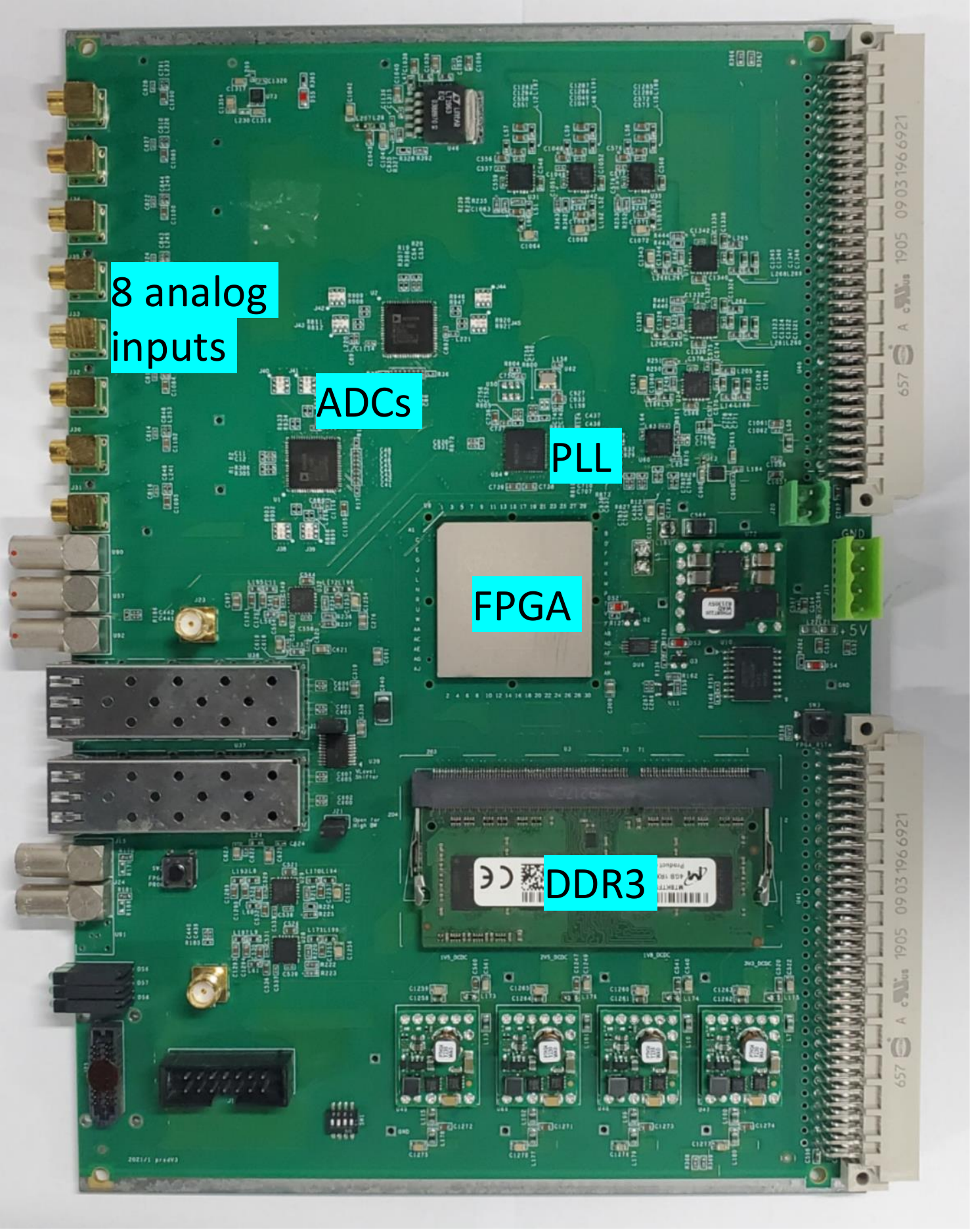}
  \caption{The schematic drawing (left) and the photo (right) of the
    new digitizer. The digitizer is a 16-layer PCB, hosting 8
    differential amplifiers (ADI LTC6409), 8 low-pass filters
    (Mini-Circuits DLFCN-290+), two ADCs (ADI AD9694), one FPGA (Xilinx XC7K325T),
    one PLL (TI LMK04610), one DDR3 memory module (Micron 4GB
    MT8KTF51264HZ-1G9P1) and other parts.  See text for more details.}
  \label{hd}
\end{figure*}

The digitizer is a 6U VME-standard module, as shown in
Figure~\ref{hd}. It combines two analog-to-digital converters (ADCs)
with a Field Programmable Gate Array (FPGA).  Each input single-ended
analog signal is converted to a differential pair through a
differential amplifier. The gain is set to be 1.6. After the
amplifier, the signal is attenuated by resistors and a low-pass filter
before entering into the ADCs. The overall amplification is about
1.25. In the ADC, four input signals are simultaneously sampled and
digitized with a sampling rate of 500 MS/s and a 14-bit resolution. The
dynamic range is set to be 2.16 Vpp. The digital data are transferred
to the FPGA through the high-speed JESD204B serialized interface~\cite{jedec}.

\begin{figure*}[!htbp]
  \centering
  \includegraphics[width=0.4\linewidth]{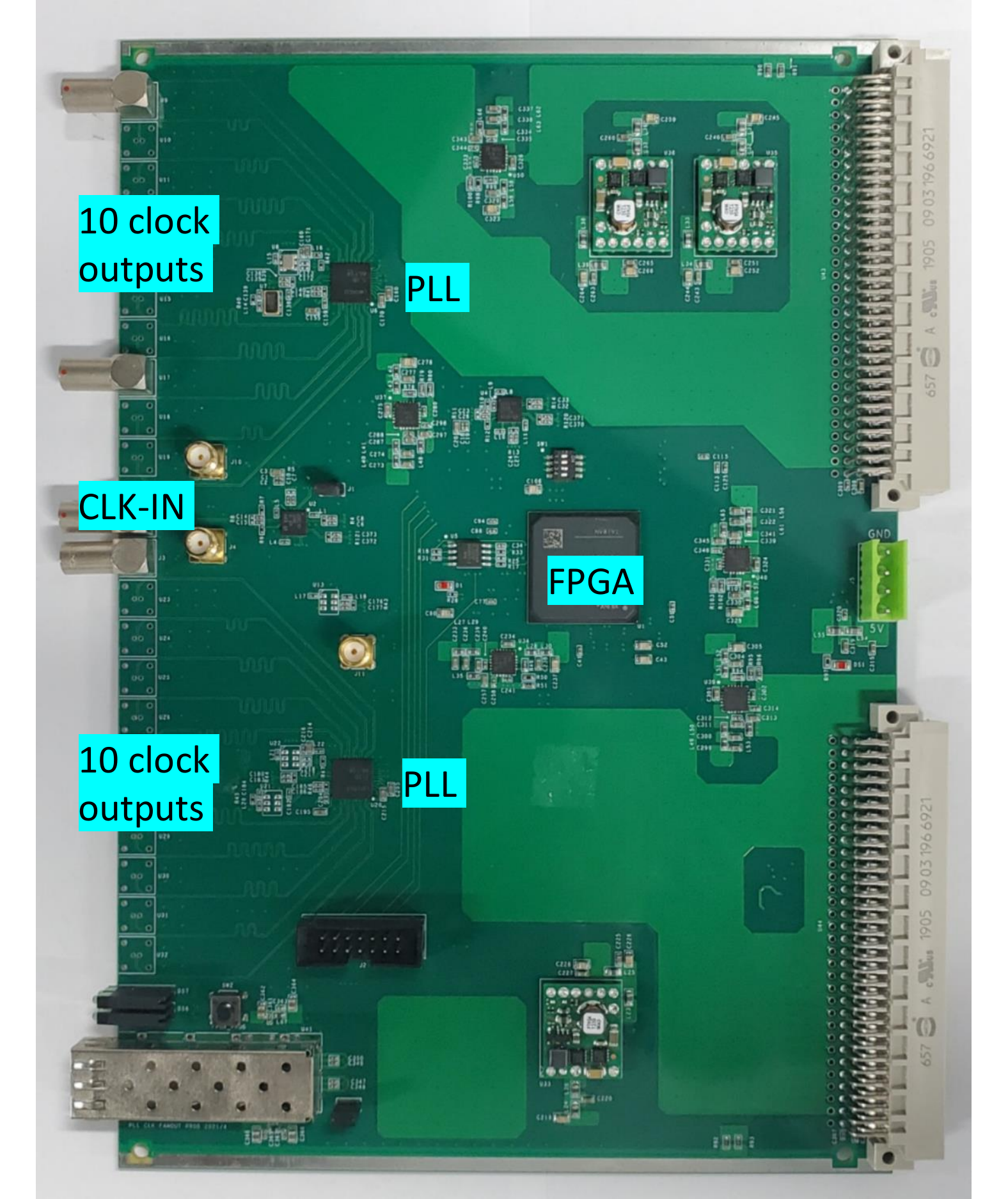}
 \caption{The clock fanout VME module, which uses the same PLLs as in
   the digitizer design. This module is a 8-layer PCB, hosting two
   PLLs (TI LMK04610), one FPGA (Xilinx XC7A35T) and other parts.
   Each module can provide at most 20 synchronous clocks with
   programmable frequencies. More outputs can be realized by chaining
   the modules.}
  \label{photo}
\end{figure*}

The ADC sampling clock and the JESD204B receiver core clock in the
FPGA are provided by a JESD204B compliant and programmable clock
jitter cleaner with internal phased-locked loops (PLLs).  The system
reference (SYSREF) signal, required by the JESD204B subclass 1
standard, acts as a common timing reference to synchronize internal
framing clocks in the ADCs and the FPGA. For synchronization among
different digitizers, a common-source external clock is required to be
the reference clock input of the PLL. This is achieved with a
self-designed clock fanout VME module (see
Figure~\ref{photo}). For each data acquisition run, common external start-run
signals (synchronized with the PLL clock) are sent into all
digitizers, so the internal trigger time counters begin to increase at
the same rising edge of the start-run signals.  In a
multi-digitizer system with the clock fanout module, the
channel-to-channel synchronization within single board and between two
boards is measured to be better than 0.2 ns, which is sufficiently
precise for PandaX DM experiments given that a typical S1 signal is
O(100) ns wide. In addition, several other clocks are required. A
clock is needed to program the ADCs and clock jitter
cleaner. Reference clocks are needed for the serial transceivers in
the FPGA (denoted as MGT\_CLK in Figure~\ref{hd} left). These clocks
are provided by a clock buffer with an oscillator as the input.
    
\begin{figure*}[!htbp]
  \centering
 \includegraphics[width=0.8\linewidth]{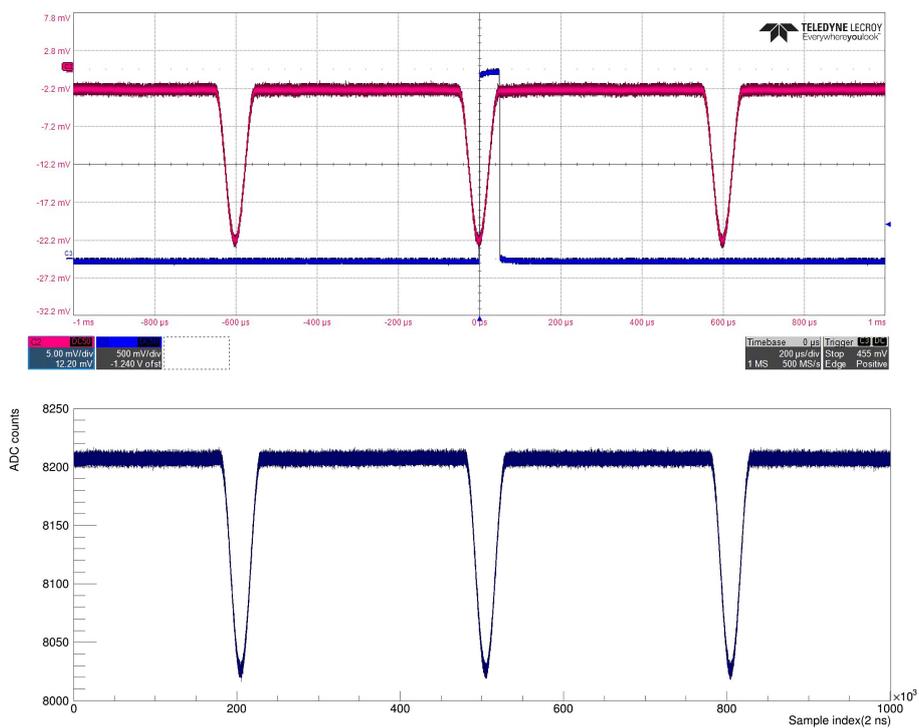}
 \caption{Top, the input signal and trigger signal recorded by an
   oscilloscope. Bottom, the waveform of the same signal recorded by
   the new digitizer operated in the external-trigger mode. This demonstrates the
   capability of recording long waveforms before and after the
   trigger.}
  \label{ddr3}
\end{figure*}

Data processing algorithms are implemented in the FPGA. For
external-trigger-based DAQ, a large time window around the trigger
time is required. In PandaX-4T commissioning runs, the maximum drift
time is 0.84 ms. The buffer space per channel is required to store at
least $\sim$1 ms of data. This alone already exceeds the internal memory
capability of the FPGA. Thus, a DDR3 SDRAM memory is used to cache the
data from ADCs. Once an external trigger is received, the data in DDR3
are read out. This is demonstrated in Figure~\ref{ddr3}, where the 2
ms-long data per channel are read out with the trigger at the middle
of the time window. This shows that the new digitizer can record both
S1 and S2 signals in one external trigger at the PandaX DM
experiments.

For self-trigger based DAQ, the real-time baseline is calculated
dynamically with a moving average algorithm. When a digitized sample
exceeds the baseline by a configurable threshold, an event header, consisting
of the board and channel number, trigger time stamp and busy time
counter, is written into a readout FIFO in the FPGA. Afterwards, the
whole pulse exceeding the threshold and a limited number of samples
before and after the pulse are saved into the same FIFO. The depth of
this FIFO is configured to be 65536 samples (13 us) for each channel. This is
large enough for our data taking described in next section,
given the self-trigger rate per channel is 200-300 Hz and the average
number of ADC samples per trigger is around 150.  However, it is
possible that the data in the readout FIFO are not read out fast enough.
When it is close to be full, the busy time counter will be increased
until the FIFO becomes available again. With this counter we can
monitor the dead time for each channel.  In addition, the self-trigger
algorithm is designed to accept external triggers.  This can
be used for LED light calibrations.

The digitizer connects to the DAQ server by a small form-factor
pluggable plus transceiver (SFP+) with a fiber optic link.  A Gigabit
Ethernet protocol SiTCP~\cite{sitcp} is used in the FPGA to transfer
data from the digitizer to the DAQ server through the SFP+. Another
SFP+ slot is reserved for future development. For example, it can be
used to send out the real-time waveforms of the digitizer.  The summed
waveforms from all digitizers can be used to generate external trigger
signals for the digitizers when operated in the external-trigger mode.
In addition, the digitizer includes several LVDS lanes
(Figure~\ref{hd} lower left region), which are reserved for future
development as well. They can provide real-time time-over-threshold
(ToT) signals, like the CAEN V1724 and V1725 digitizers. Such ToT
signals were used for the trigger system in previous experiments
PandaX-I~\cite{Ren:2016ium} and PandaX-II~\cite{Wu:2017cjl}. Unlike
the TOT signals, the summed waveforms carry the absolute pulse height
information. We might be able to improve the trigger performance with the new digitizers
using the summed waveforms.
 
\section{Performance}

\subsection{Performance in PandaX-4T}

\begin{figure*}[!htbp]
  \centering
  \includegraphics[width=0.8\linewidth]{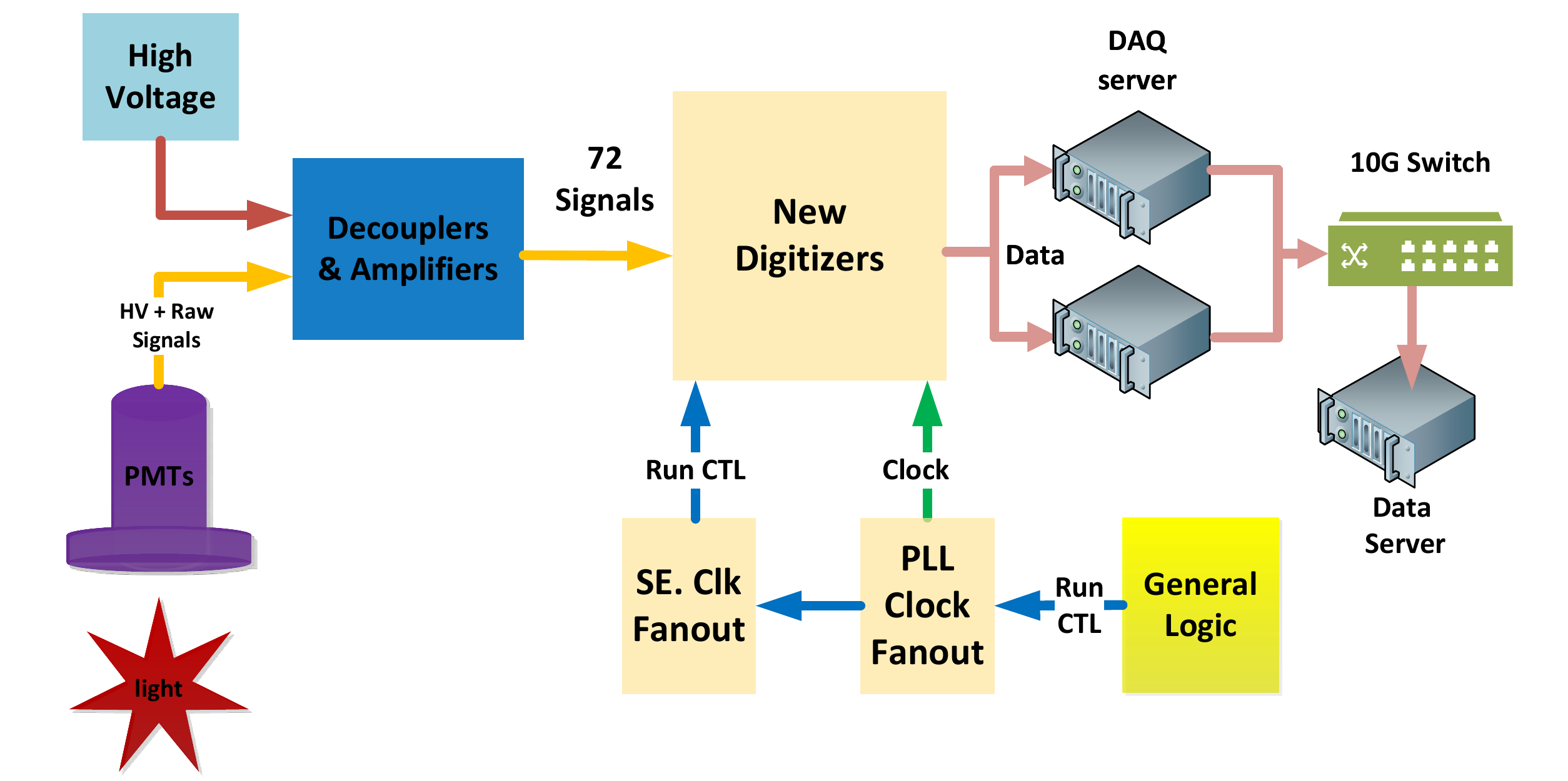}
  \includegraphics[width=0.6\linewidth]{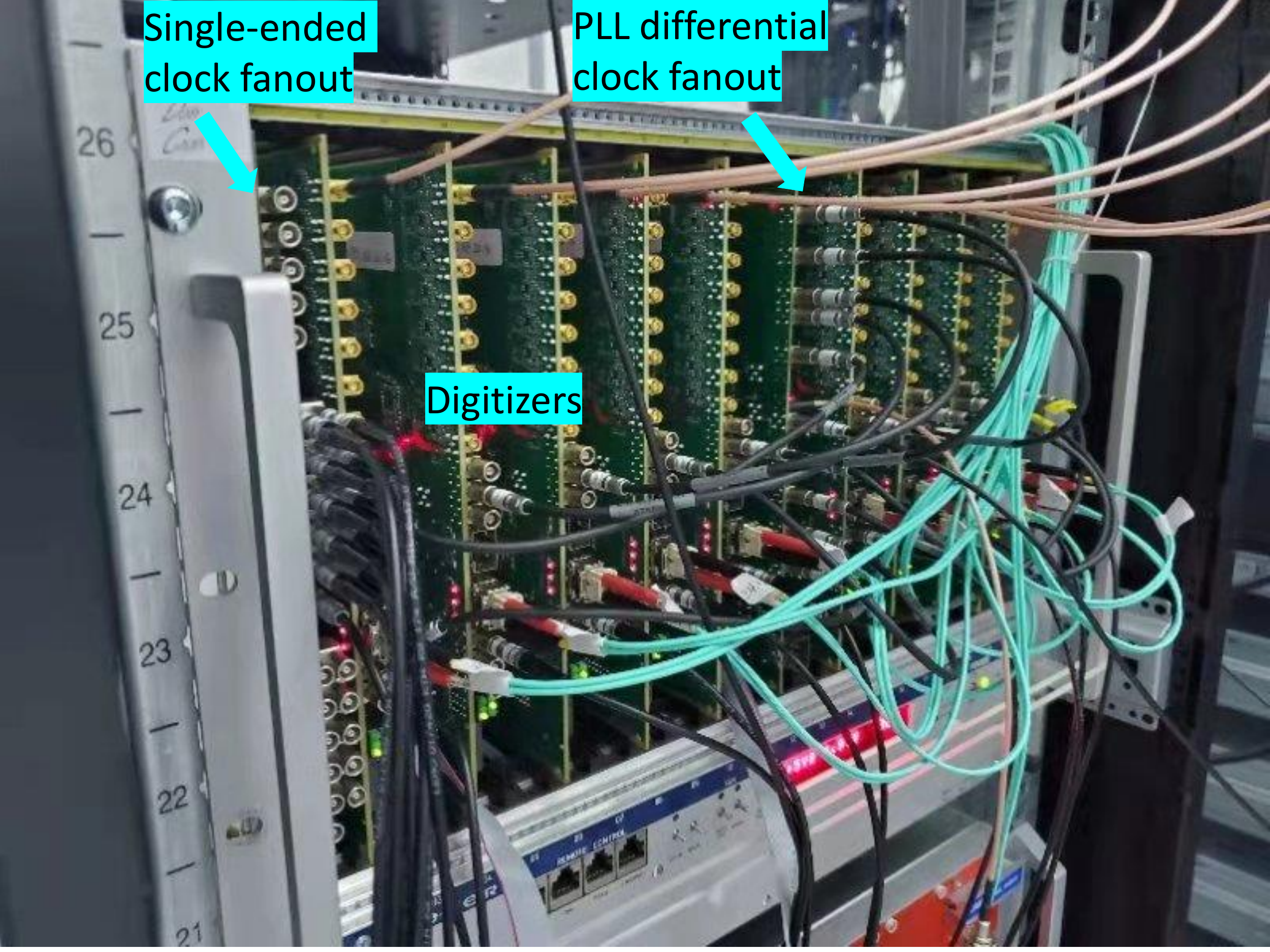}
  \caption{Top, the schematic drawing of the 72-channel new digitizer
    and DAQ system. Bottom, a photo of the new digitizer system in
    situ of PandaX-4T. This system follows the overall design of the
    current system in PandaX-4T~\cite{p4daq}. The digitizers operate
    with common clock signals distributed by the PLL-based
    differential-ended clock fanout module (Figure~\ref{photo}).  The
    start-run signals of all digitizers come from a pulse originated
    from a general logic unit (CAEN V1495). This signal is
    synchronized with the clock in the PLL clock fanout module and
    distributed to each digitizer by a single-ended clock fanout
    module used in PandaX-4T~\cite{p4daq}. }
  \label{vme}
\end{figure*}

To demonstrate the performance of the new digitizers, we built a
72-channel digitizer and DAQ system (see Figure~\ref{vme}) in situ of
PandaX-4T experiment in June 2021, after its commissioning runs. 72
bottom 3-inch PMTs (Hamamatsu R11410-23) are used for the data taking.
The DAQ system is similar as the current one in
PandaX-4T~\cite{p4daq}.  The data of each digitizer are transmitted to
a DAQ server through individual fiber optic links with commercial
PCI-E Ethernet cards. Two DAQ servers are used.  All data are
transferred to another data server and saved into the disk for offline
analysis.

\begin{figure*}[!htbp]
  \centering
  \includegraphics[width=0.45\linewidth]{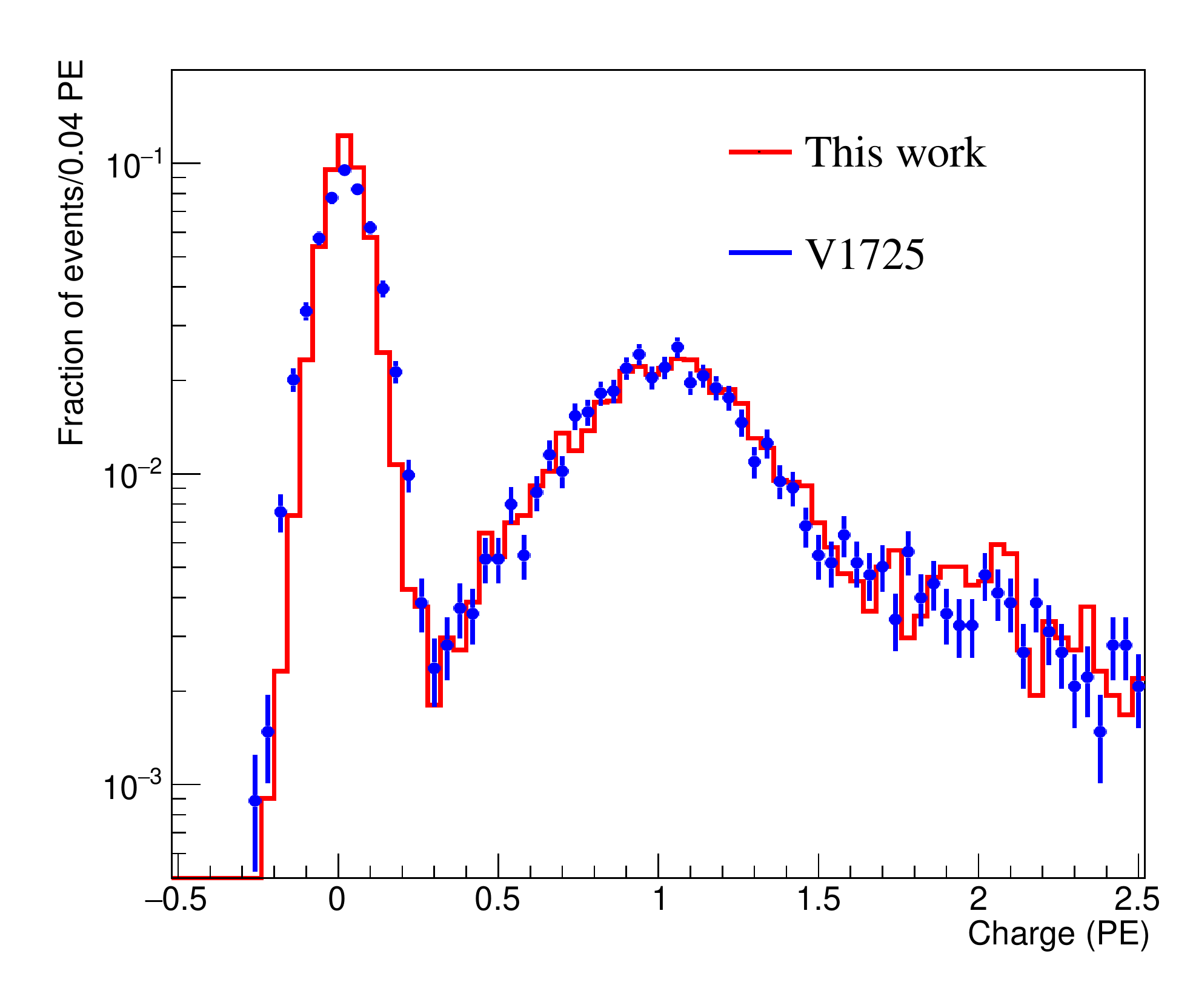}
  \includegraphics[width=0.45\linewidth]{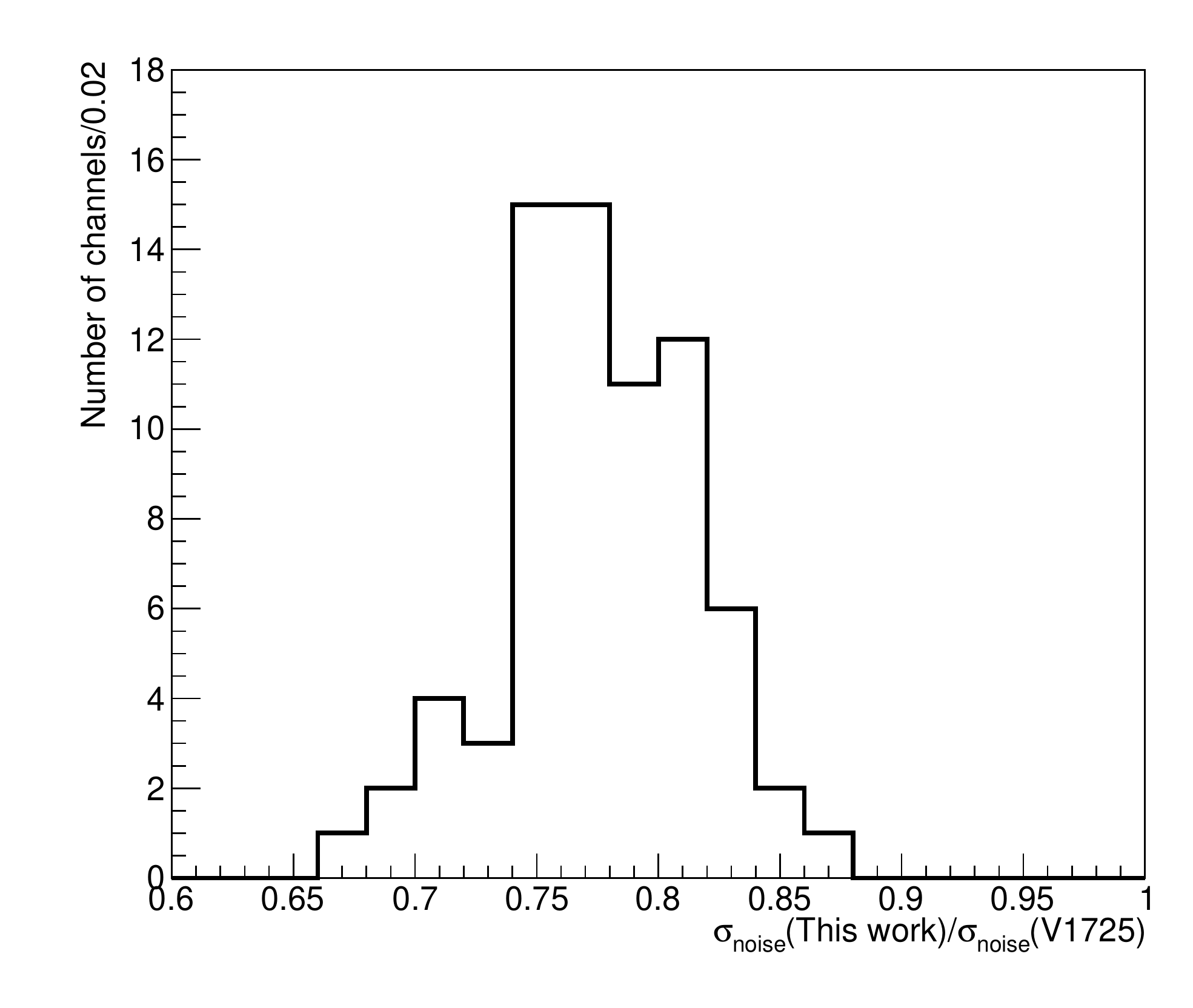}
  \caption{Left, charge distributions of a 3-inch PMT (Hamamatsu
    R11410-23) using a low-intensity LED, measured with the new digitizer (red
    curve) and the V1725 digitizer (blue dot). Right, the ratio of the pedestal width
    measured with the new digitizers and V1725s, from 72 3-inch PMTs.}
  \label{spe}
\end{figure*}

The system performance is evaluated in two folds. First is the
measurement of SPE signals, which are the smallest signals in the
experiment.  For the new system, the gain of each PMT
channel (including the digitizer) is recalibrated with SPE signals
using a low-intensity LED. The charge distribution with pedestal contribution
of a typical 3-inch PMT is shown in Figure~\ref{spe} left. Compared to
the V1725 digitizers, the SPE distribution is almost unchanged when
using the new digitizers, because it is dominated by the PMT gain
fluctuations. However, due to the higher sampling rate and the
amplifiers in the new digitizers, the pedestal width is reduced from
0.093 PE to 0.072 PE on average. This is equivalent to a 30\% improvement
on the signal-over-noise ratio.

\begin{figure*}[!htbp]
  \centering
  \includegraphics[width=0.45\linewidth]{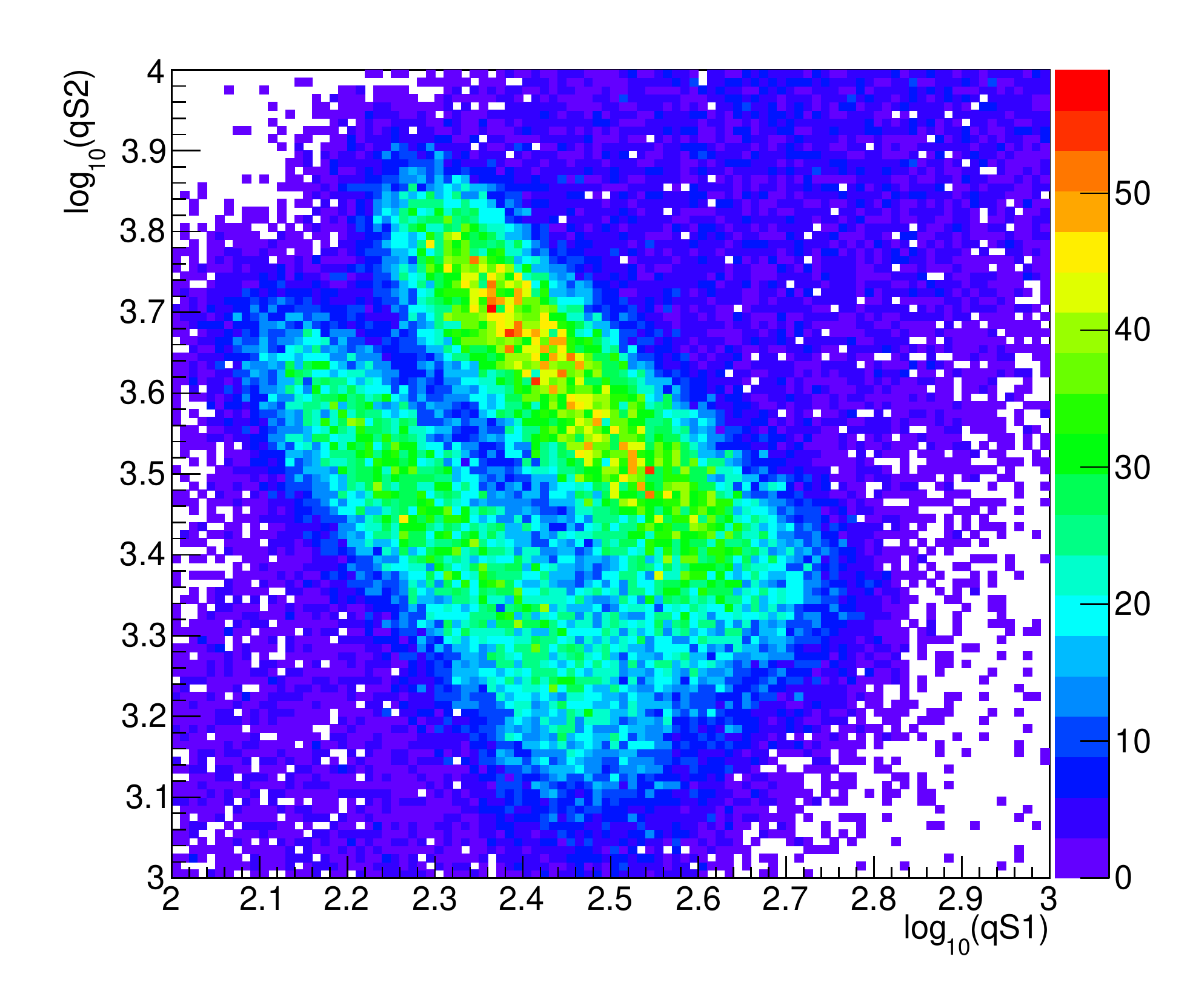}
  \includegraphics[width=0.45\linewidth]{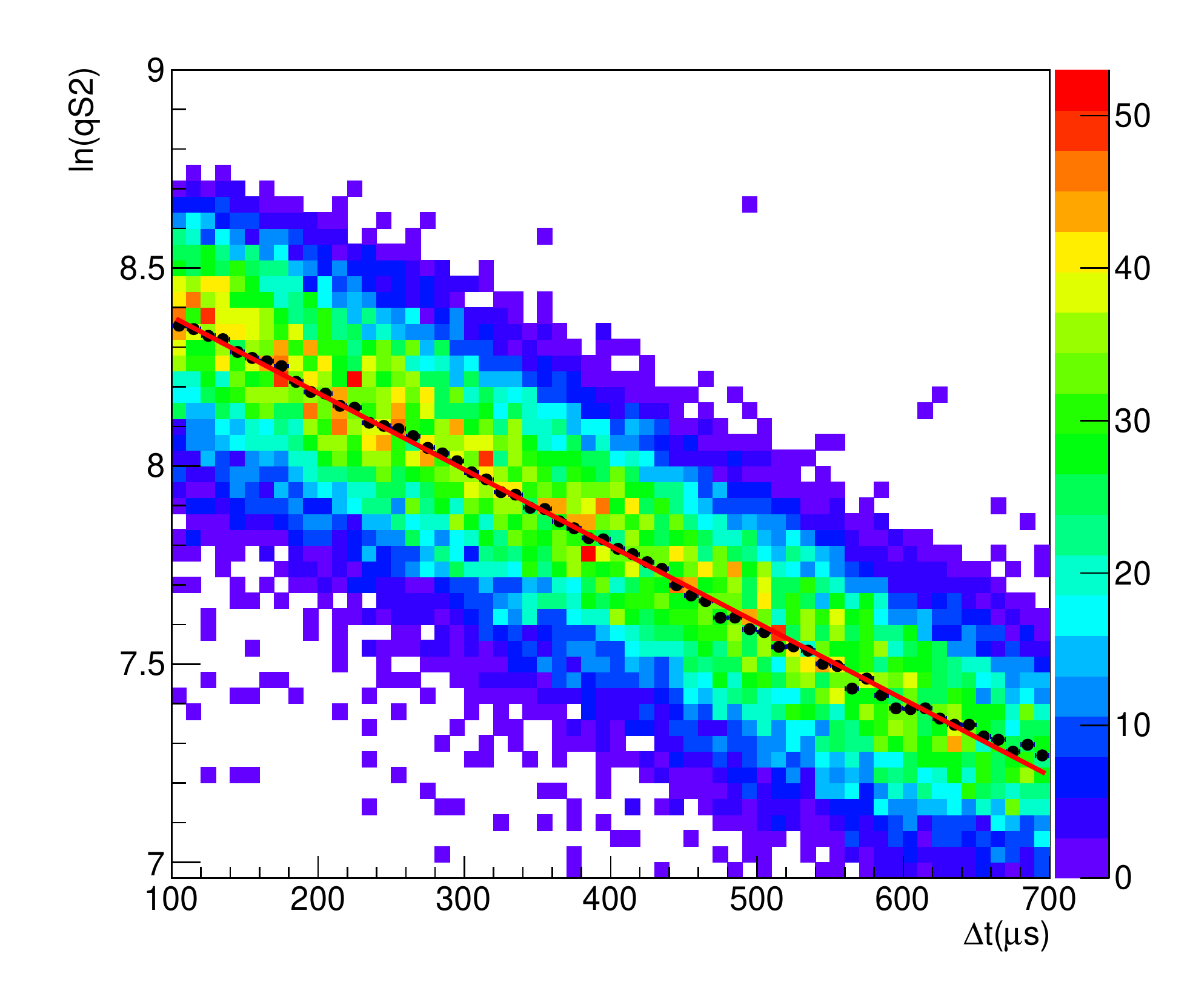}
  \includegraphics[width=0.45\linewidth]{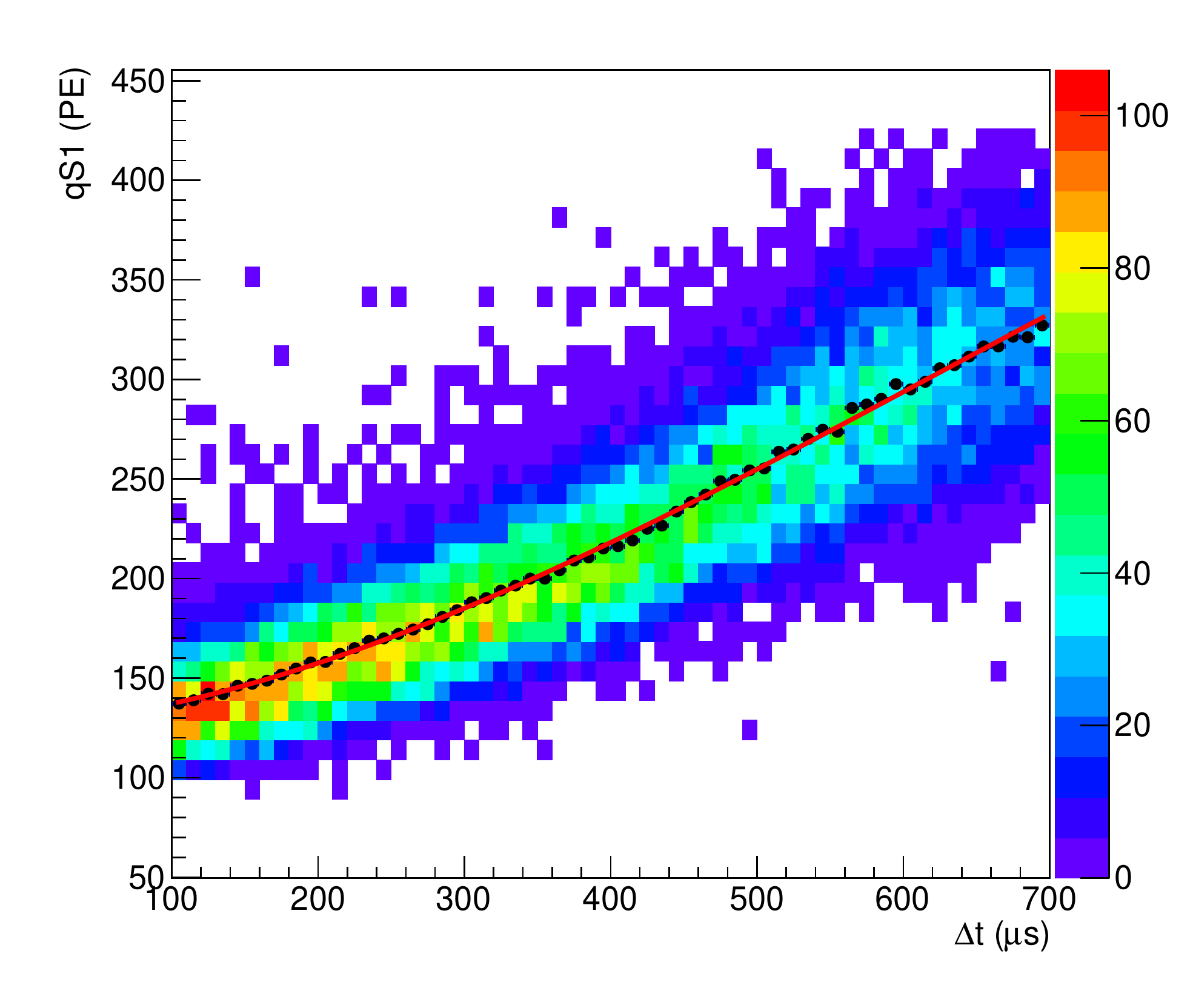}
  \includegraphics[width=0.45\linewidth]{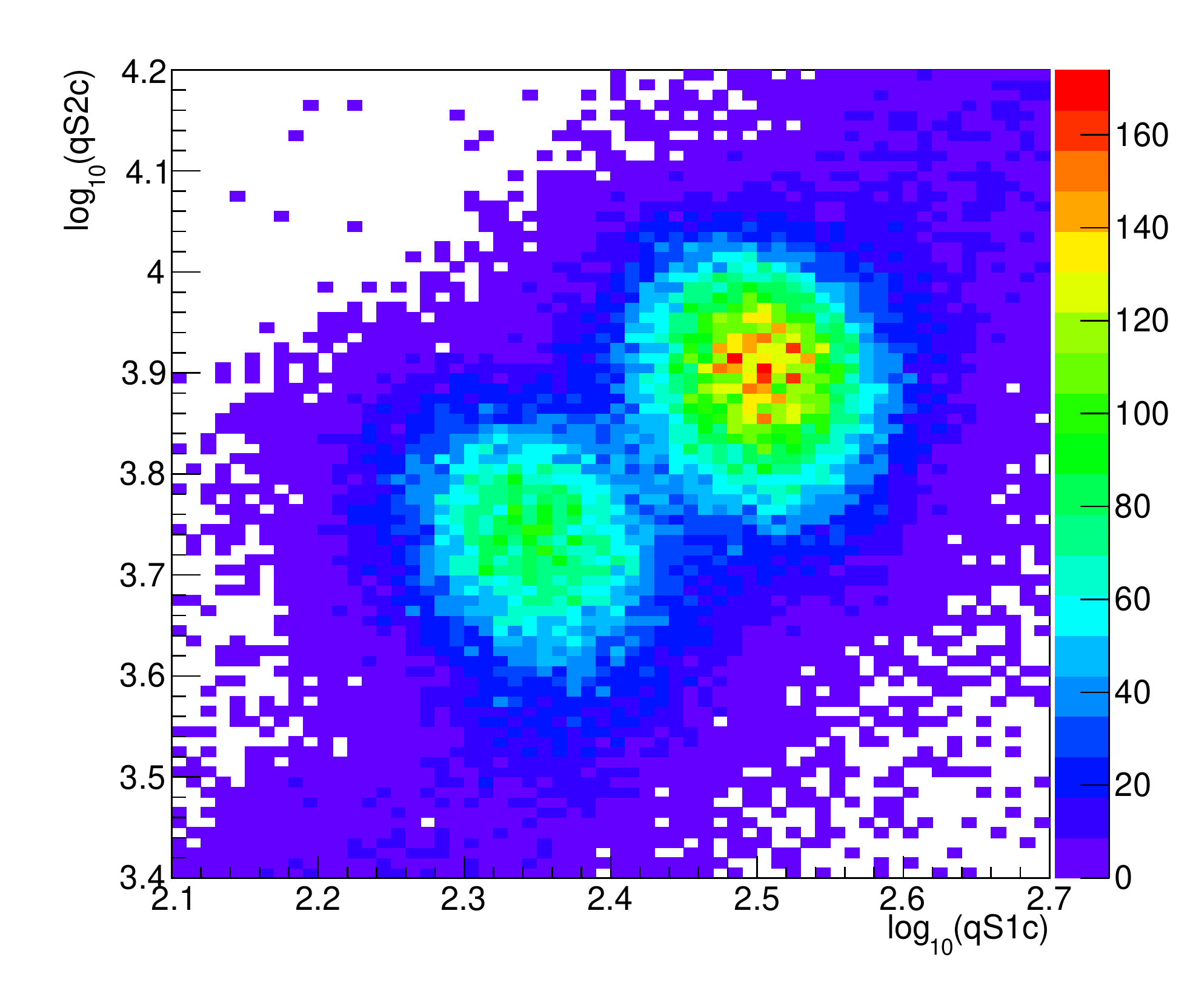}
  \caption{Top left, the uncorrected charge of S2 vs S1 signals
    measured with the 72-channel new digitizer system.  The two bright bands
    correspond to the 164 keV and 236 keV $\gamma$s from $^{131m}$Xe
    and $^{129m}$Xe decays, respectively. Top right, the dependence of
    S2 uncorrected charge on the electron drift time. The data points
    are the average charge per drift time bin. Superimposed is an
    exponential function fitted to the data.  Bottom left, the
    dependence of S1 uncorrected charge on the drift time. The curve
    is a 3rd-order polynomial function fitted to the data. Bottom
    right, the corrected charge of S2 vs S1 signals.}
  \label{qS}
\end{figure*}

Second is the reconstruction of physical events. A 14.8-hour dataset
were recorded using the new digitizers, operated in the triggerless mode.
The self-trigger threshold is set to be 22 ADC counts. For a
typical 3-inch PMT channel, this corresponds to 0.28 PE. In this run,
no dead time is observed for most channels. For others, the dead time
is negligible. The maximum dead time among all channels is 0.003 ms
during the 14.8 hours of data taking.  In the same run, data from
other PMT channels were recorded using V1725 digitizers. The typical
threshold is about 1/3 PE~\cite{p4daq}. Before this run, the liquid
xenon was irradiated for 1/2 day with a Pu-C neutron source.  So we
expect to record a large number of high-energy $\gamma$ events with energies around 164
keV ($^{131m}$Xe) and 236 keV ($^{129m}$Xe). The event rates of the
two $\gamma$ rays measured from two independent digitizer systems are
expected to be consistent.

The rates are measured from the reconstructed energy distribution. The
procedures are as follows.  The recorded raw data are
processed in a similar way as the first PandaX-4T DM
analysis~\cite{p4paper}. Hits with amplitudes above the trigger
threshold are identified from individual channels. Signal is then
defined as a cluster of hits with tail-to-head gap not greater than 60
ns, in order to capture nearly all scintillation lights from xenon
excimers with lifetime $\sim$4 ns and $\sim$21 ns from the single and
triplet states, respectively.  The inefficiency of such clustering gap
requirement has been studied in Ref.~\cite{p4paper} and found to be
negligible using data driven approaches. The S1-like and S2-like
signals are tagged based on the width of the cluster waveform
enclosing 10\% to 90\% cumulative charge. In addition, S1 and S2 signals are required to have a charge qS1$>$5 PE and qS2$>$200 PE, respectively.

An event is then defined by combining all S1-like and S2-like signals
within a time window of 1 ms. The S1 and S2 signals with largest
charges are selected to form a pair. The electron drift time ($\Delta t$
between S2 and S1 signals) is required to be above 100 $\mu$s and
less than 700 $\mu$s. This serves as a fiducial volume selection which
removes $\sim$68\% background events, while keeping $\sim$70\% signals events from the two
high-energy $\gamma$s. No other selection cuts are applied. The correlation between
S2 and S1 charge is shown in the Figure~\ref{qS} top left panel. There
are two distinct bands. The lower and the upper one correspond to the
164 keV and the 236 keV $\gamma$s, respectively.

The qS2 and qS1 are corrected before they are used to reconstruct
the energy.  Due to electronegative impurities in the liquid, qS2 is
exponentially reduced as the drift time increases, as shown in the
Figure~\ref{qS} top right panel. Here, the 164 keV data are selected
around the lower band in the Figure~\ref{qS} top left panel. The corrected S2 charge is given by
qS2c$=$qS2$\times e^{\Delta t/\tau}$, where $\tau$ is the electron life time obtained from
the fit in Figure~\ref{qS} top right panel. When the interaction happens closer to the liquid surface, the scintillation lights of S1 signals are more difficult to be detected by the PMTs due
to light reflections on the TPC wall. This is shown in the Figure~\ref{qS} bottom
left panel using the same 164 keV data. The corrected S1 charge becomes qS1c$=$qS1$/f(\Delta t)\times<f(\Delta t)>$, where $f(\Delta t)$ is the polynomial function in Figure~\ref{qS} bottom left panel, and $<f(\Delta t)>$ represents the average value of the fucntion from 100 and 700 $\mu$s. The corrected charge qS2c vs qS1c is shown
in the Figure~\ref{qS} bottom right panel. As expected, the two bands become more localized
after corrections. 

\begin{figure*}[!htbp]
  \centering
  \includegraphics[width=0.45\linewidth]{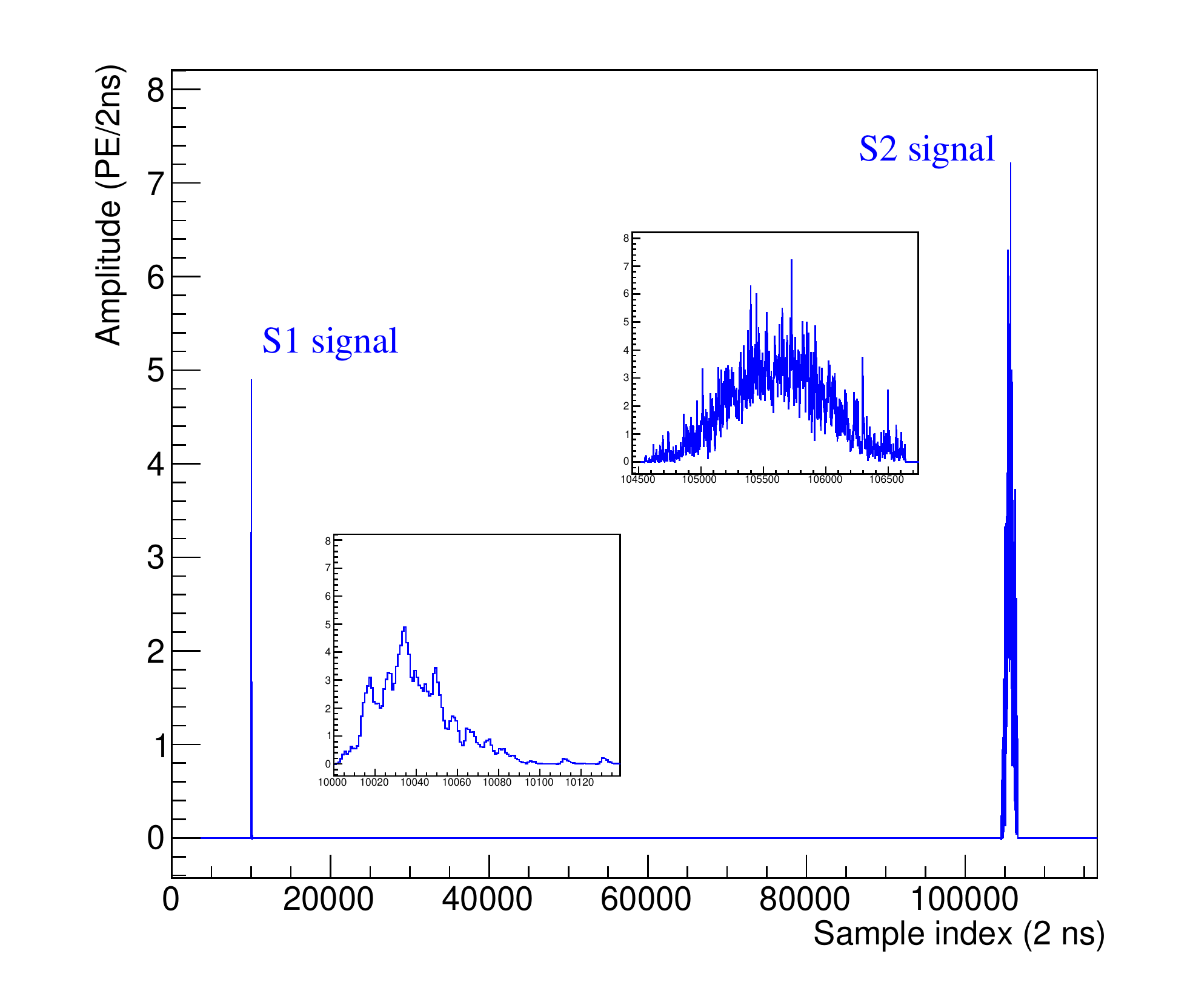}
  \includegraphics[width=0.45\linewidth]{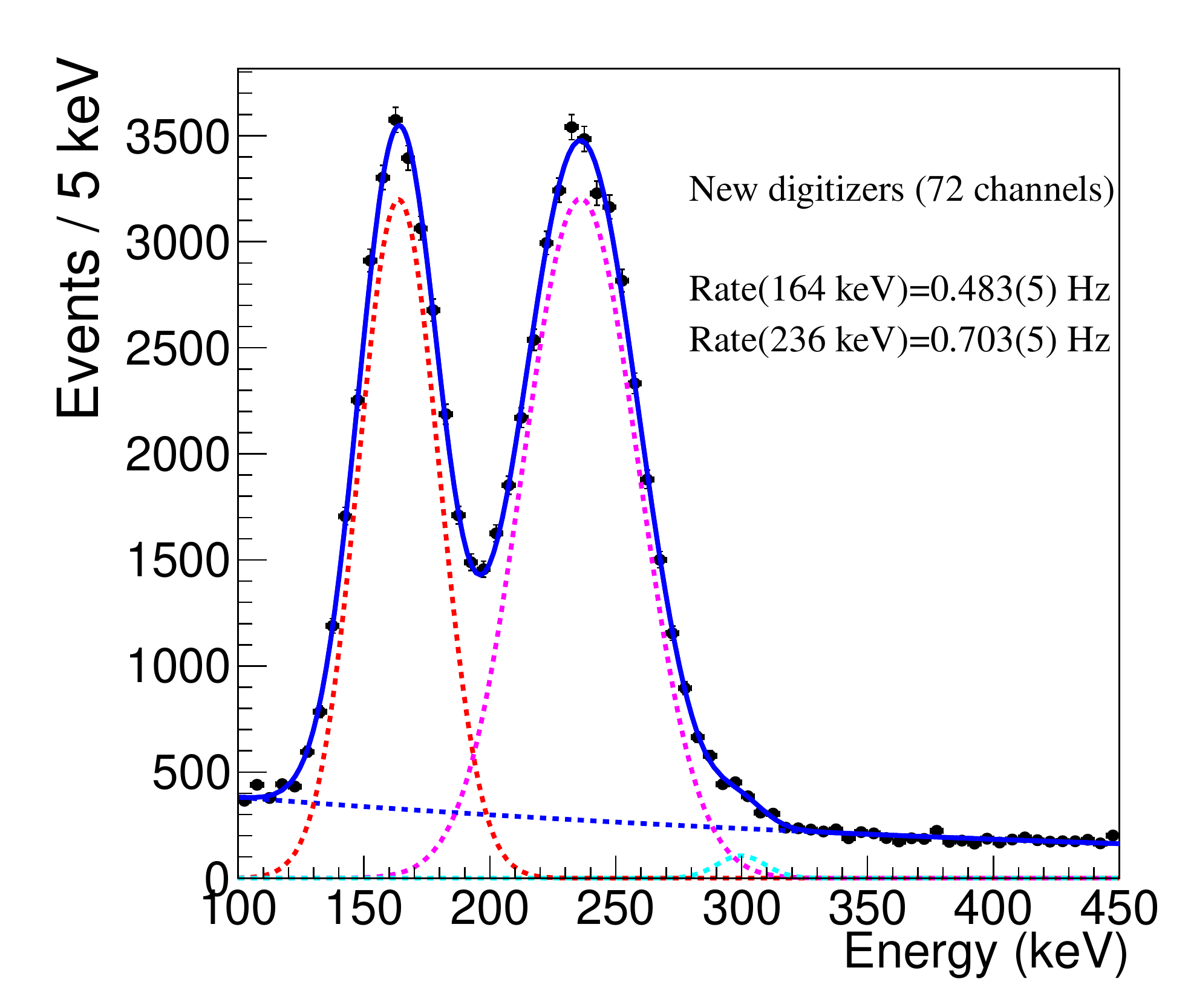}
  \caption{Left, typical waveforms of the S1 and S2 signals from a 164
    keV $\gamma$ ray, reconstructed with the 72-channel new digitizer
    system.  Right, the reconstructed energy distribution, which is
    fitted with the sum (blue solid line) of three Gaussian functions and an exponential
    function (blue dotted line). The first two Gaussian functions describe the 164 keV (red dotted line)
    and 236 keV (magenta dotted line) main peaks, while the last one describes the
    non-gaussian tail of the 236 keV peak (cyan dotted line). Similar tail also exists in the distribution reconstructed using V1725 digitizers. The same form of functions is used to fit the data distribution and extract the signal yields for comparison.}
  \label{energy}
\end{figure*}

Finally, the combined energy is reconstructed as E$=13.7 $eV $\times (\mathrm{qS1c}/g1+\mathrm{qS2c}/g2)$, where $g1$ is the photon detection efficiency, $g2$ is the product of the electron extraction
efficiency and the single electron gain.  The $g2/g1$ is taken from
Ref.~\cite{p4paper} but adjusted taking into account that only bottom
PMTs are used here for the S1 charge measurement while both top and
bottom PMTs are used in Ref.~\cite{p4paper}. The values of $g1$ and $g2$
are solved such that the 164 keV distribution is peaked at the right position.
Figure~\ref{energy} left shows a typical waveform of the S1 and S2 signals with charges and combined energy
compatible with a 164 keV deposition in the liquid.  Figure~\ref{energy} right shows the energy distribution. An extended likelihood fit is used to extract the number of signal events. The event rates for the 164 keV and 236 keV $\gamma$ rays are
measured to be $0.483\pm0.005$ Hz and $0.703\pm0.005$ Hz, respectively.

The data recorded with V1725 digitizers in the same run are processed
with the same procedures described above. The $g2/g1$ is also taken
from Ref.~\cite{p4paper} and adjusted according to the fact that 72 bottom
PMTs are not used for S1 and S2 measurements. The measured event rates
of the two $\gamma$ rays are $0.494\pm0.003$ Hz and $0.690\pm0.005$
Hz, respectively. These results are consistent with the previous
measurements within two standard deviations. This provides a cross
validation for our new digitizers and the V1725 digitizers.

\subsection{Performance with the new PMTs for PandaX-30T}
For the future experiment PandaX-30T, a new type of 2-inch
PMTs (Hamamatsu R12699-406-M4) are being considered. Compared to the
3-inch PMTs used in PandaX-4T experiment, the new PMTs have smaller
radioactivity level. In addition, each PMT consists of 4
$1\times1$-inch PMT array. So these PMTs can provide better photon
coverage and better position measurement in dual-phase xenon TPC
experiments. 

\begin{figure*}[!htbp]
  \centering
  \includegraphics[width=0.6\linewidth]{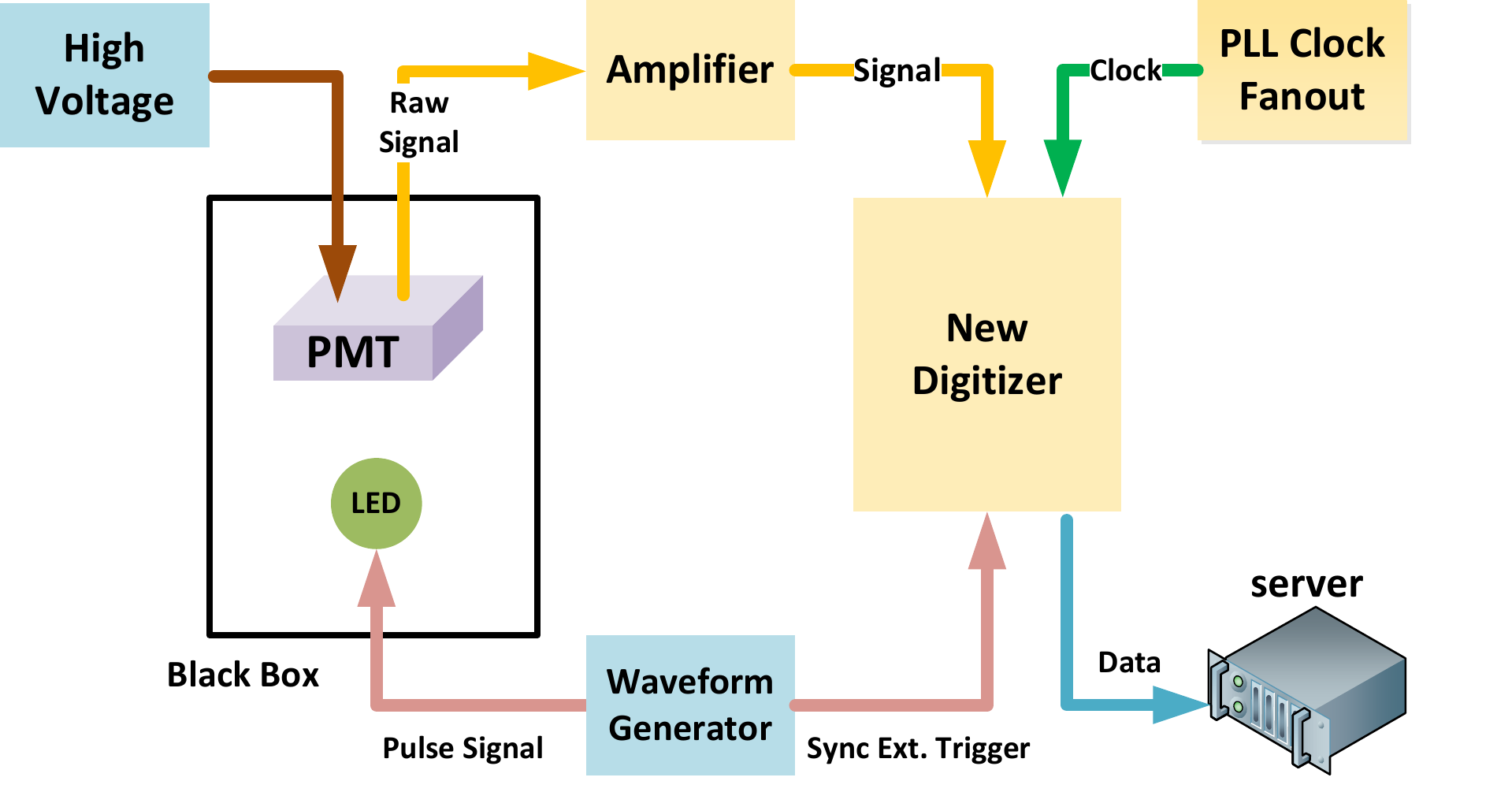}
  \caption{The schematic drawing of the setup used to measure the SPE signals of the new 2-inch PMTs
    (Hamamatsu R12699-406-M4). A high voltage of
    1000 V is applied to the PMT base board. Four raw signals of each
    PMT are decoupled from the HV on the base board and get amplified
    by a factor of four before entering into the new digitizer. The amplification
    circuit board consists of high-speed current feedback amplifiers (ADI AD8009\cite{adi}).
    The waveform generator sends a 100 Hz pulse to drive the LED to emit
    lights. A synchronous trigger is sent to the digitizer to record the
    data, which are sent to a server for offline analysis.}
  \label{setup}
\end{figure*}

\begin{figure*}[!htbp]
  \centering \includegraphics[width=0.45\linewidth]{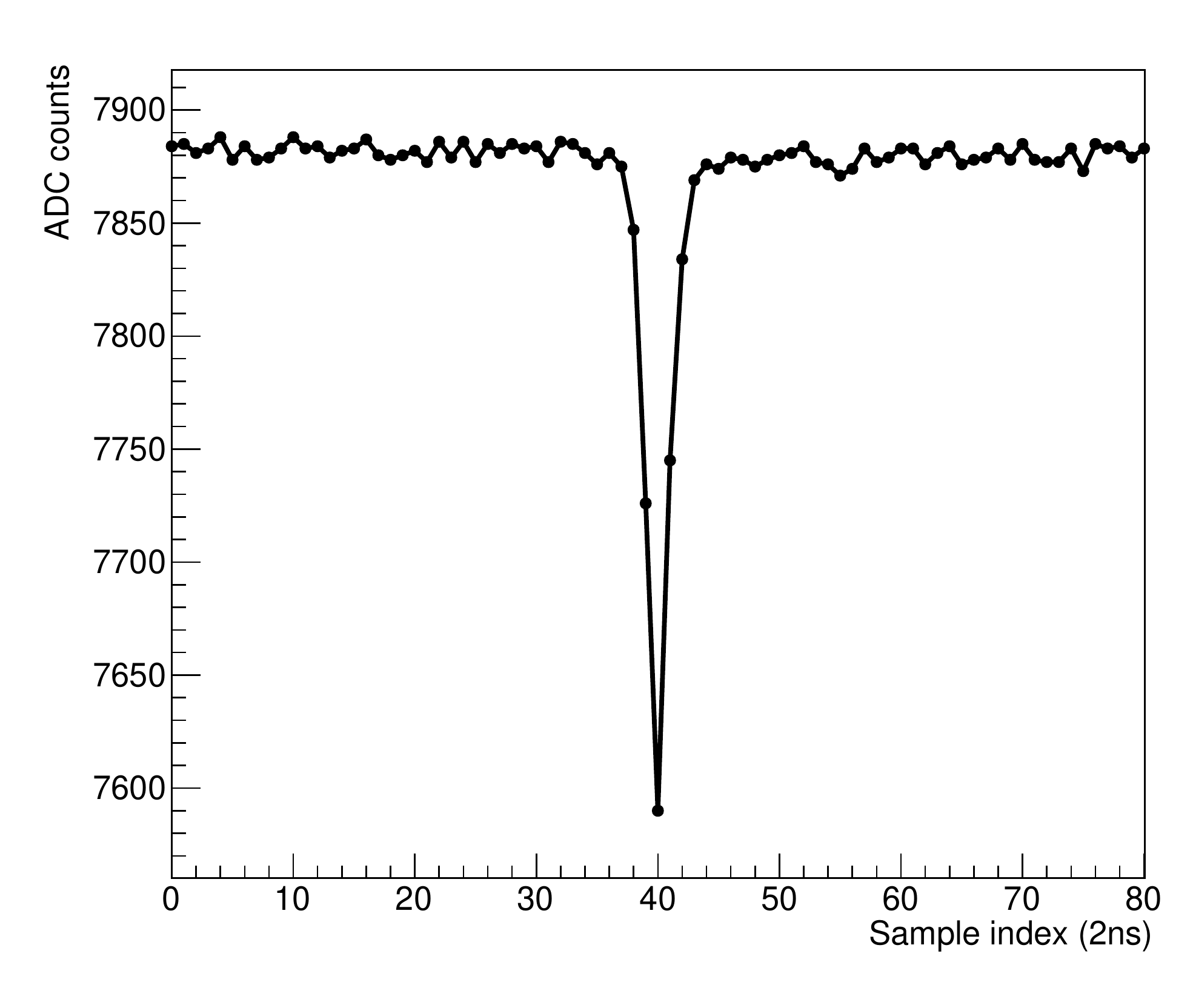}
  \includegraphics[width=0.45\linewidth]{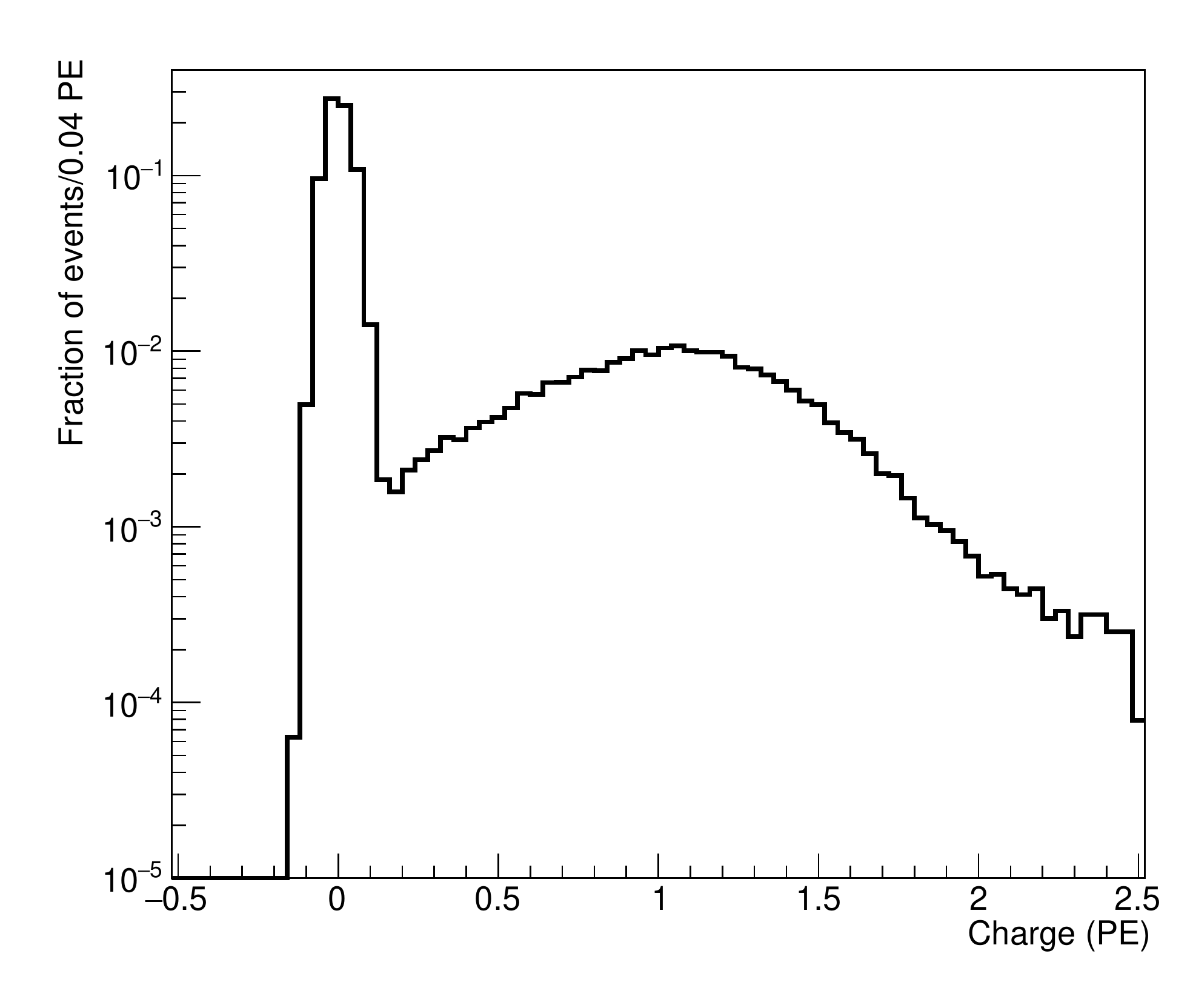}
  \caption{Left, a typical SPE waveform from a 2-inch PMT (Hamamatsu
    R12699-406-M4) recorded by our new digitizer using a low-intensity
    LED. Right, the charge distribution from one output channel of the
    2-inch PMT.}
  \label{spe_new}
\end{figure*}

Our new digitizers have been used to evaluate the new PMTs, with the setup shown in Figure~\ref{setup}.
Figure~\ref{spe_new} left shows a typical waveform of the SPE signals
recorded by the new digitizer. The typical SPE signal pulse lasts
around 10-15 ns, which is about a factor two shorter than the 3-inch PMTs
used in PandaX-4T. Given that a 250 MS/s digitizer is used in
PandaX-4T, a 500 MS/s sampling rate is a reasonable choice for the new PMTs.
Figure~\ref{spe_new} right shows the measured charge
distribution from one output channel of the new PMT. The SPE peak is clearly distinguishable
from the pedestal.  

\section{Summary}
In summary, we presented a 8-channel 500 MS/s waveform digitizer that is
designed to support both triggerless readout and external-trigger-based readout. Using real data from PandaX-4T, the triggerless readout
of the new digitizers have been cross validated with the current
digitizers used in the experiment. The new digitizers have been used
to evaluate the new 2-inch PMTs for future experiment. The
capability of recording long waveforms before and after the external
trigger is demonstrated with a waveform generator. This shows the new
digitizer should be able to record both S1 and S2 signals in one
trigger time window. A trigger system is under development. The system
is expected to analyze the real-time waveforms from the new digitizers
and provide external triggers for the digitizers when operated in the
external-trigger mode. A joint test of the trigger system and the
digitizer system will be performed in the future.

\section{Acknowledgement}
This project is supported by grants from the Ministry of Science and Technology
of China (No. 2016YFA0400301 and 2016YFA0400302), a Double Top-class grant from
Shanghai Jiao Tong University, grants from National Science Foundation of China
(Nos. 11875190, 11505112, 11775142, 11755001 and 12090063), supports from the Office of
Science and Technology, Shanghai Municipal Government (18JC1410200),
and support also from the Key Laboratory for Particle Physics,
Astrophysics and Cosmology, Ministry of Education. This work is
supported also by the Chinese Academy of Sciences Center for
Excellence in Particle Physics (CCEPP). We thank the PandaX-4T
collaboration for their support for our in situ test of the new
digitizers. We acknowledge useful discussions and help on the hardware
development and test from Qi An, Shubin Liu, Changqing Feng, Zhongtao Shen,
Shuwen Wang at the State Key Laboratory of Particle Detection and
Electronics, University of Science and Technology of China.

\end{document}